\documentclass[twocolumn,showpacs,preprintnumbers,amsmath,amssymb]{revtex4-1}

\usepackage{graphicx}
\usepackage{dcolumn}
\usepackage{bm}
\usepackage{color}

\begin{document}

\preprint{}

\title{
Sfermion loop contribution to the two-loop level fermion electric dipole moment
in $R$-parity violating supersymmetric models}

\author{Nodoka Yamanaka}
\affiliation{%
Research Center for Nuclear Physics, Osaka University, Ibaraki, Osaka 567-0047, Japan}%



\date{\today}

\begin{abstract}
We evaluate the Barr-Zee type two-loop level contribution to the fermion electric and chromo-electric dipole moments with sfermion loop in $R$-parity violating supersymmetric models.
It is found that the Barr-Zee type fermion dipole moment with sfermion loop acts destructively to the currently known fermion loop contribution, and that it has small effect when the mass of squarks or charged sleptons in the loop is larger than or comparable to that of the sneutrinos, but cannot be neglected if the sneutrinos are much heavier than loop sfermions.

\end{abstract}

\maketitle

The supersymmetric extension of the standard model is one of the favored candidates of new physics. 
It has many advantages which resolve theoretical and observational problems of the standard model (SM), and has been exhaustively studied over decades \cite{mssm}.
The supersymmetric SM allows also baryon number or lepton number violating 
interactions, known as the $R$-parity violating (RPV) interactions.
The RPV effects have also been constrained from the analysis of various phenomena \cite{rpvphenomenology}. 

In searching for new physics beyond the SM, the electric dipole moment (EDM) plays a very important role.
The EDM is an observable sensitive to the P and CP violations of the underlying theory, and provides excellent information about the new physics for many reasons.
First, it can be measured in many systems and many accurate experimental data are available \cite{edmexp}.
Second, it has a very small SM contribution \cite{smedm}, so that any observation of finite EDM will be a direct indication of new physics.
Finally, the EDM is sensitive to many new physics, in particular the supersymmetry, and many supersymmetric models with \cite{susyedm1-loop,susyedm2-loop,susyedmgeneral,susyedmflavorchange} and without 
\cite{barbieri,godbole,herczege-n,chang,choi,faessler,yamanaka1,yamanaka2} the conservation 
of $R$-parity have been studied.

In the RPV supersymmetric model with trilinear RPV interactions, it has been
found that the fermion 
(quark or lepton) EDM does not receive any one-loop level contribution \cite{godbole}, and 
the two-loop level contribution has been analyzed in detail to give the Barr-Zee type diagram \cite{barr-zee} with fermion loops as the leading contribution \cite{chang,yamanaka1}.
In previous works however, Barr-Zee type process with sfermion loops has not been considered so far, although it involves the same set of coupling constants.
In this paper we will evaluate the Barr-Zee type diagrams with sfermion loop generated by RPV interactions to clarify the relative size between the currently known fermion loop contribution.
It is actually found that the sfermion loop effect acts destructively to the fermion loop one, and that it can be significant in some situation.

The RPV interactions are generated by the following superpotential:
\begin{eqnarray}
W &=&
\frac{1}{2} \lambda_{ijk} \epsilon_{ab} L_i^a L_j^b (E^c)_k
+\lambda'_{ijk} \epsilon_{ab} L_i^a Q_j^b ( D^c)_k \nonumber\\
&&
+(Y_{u})_{ij} \epsilon_{ab} Q_i^a H_u^b U_j^c
+(Y_{d})_{ij} Q_i^a H_{da} D_j^c \nonumber\\
&&\hspace{2em}+(Y_{e})_i L_i^a H_{da} E_i^c
\ .
\label{eq:rpvsuperpotential}
\end{eqnarray}
In the first line are written RPV terms, and in the last two lines the standard $R$-parity conserving Higgs-matter interactions.
Indices $i,j,k=1,2,3$ indicate the generation, and $a,b=1,2$ the $SU(2)_L$ indices. 
$L$ and $E^c$ denote the lepton doublet and singlet left-chiral superfields. $Q$, $U^c$ and $D^c$ denote respectively the quark doublet, up quark singlet and down quark singlet left-chiral superfields.
The Higgs fields are denoted by $H_u$ and $H_d$.
The RPV baryon number violating interactions are irrelevant in this analysis since they do not contribute to the Barr-Zee type diagrams,
and are not included in our current analysis.
Also the bilinear RPV interactions were not considered.

The lagrangian is derived with the following formula:
\begin{equation}
{\cal L} = 
-\frac{1}{2} \sum_{i,j} \bar \psi_i \frac{\partial^2 W }{ \partial \phi_i \partial \phi_j} \psi_j + {\rm h.c.} 
-\sum_i \left| \frac{\partial W}{\partial \phi_i} \right|^2 \, .
\end{equation}
We then obtain the following RPV lagrangian:
\begin{widetext}
\begin{eqnarray}
{\cal L }_{\rm R\hspace{-.5em}/\,} &=&
- \frac{1}{2} \lambda_{ijk} \left[
\tilde \nu_i \bar e_k P_L e_j +\tilde e_{Lj} \bar e_k P_L \nu_i + \tilde e_{Rk}^\dagger \bar \nu_i^c P_L e_j -(i \leftrightarrow j ) \right] + ({\rm h.c.})\nonumber\\
&&-\lambda'_{ijk} \left[
\tilde \nu_i \bar d_k P_L d_j + \tilde d_{Lj} \bar d_k P_L \nu_i +\tilde d_{Rk}^\dagger \bar \nu_i^c P_L d_j -\tilde e_{Li} \bar d_k P_L u_j - \tilde u_{Lj} \bar d_k P_L e_i - \tilde d_{Rk}^\dagger \bar e_i^c P_L u_j \right]  + ({\rm h.c.}) \nonumber\\
&&
- \frac{1}{2} \lambda_{ijk} \left[ 
( m_{e_j} \tilde \nu_i \, \tilde e_{Rj} - m_{e_i} \tilde \nu_j \, \tilde e_{Ri})\, \tilde e_{Rk}^\dagger
+m_{e_k} (\tilde \nu_i \, \tilde e_{Lj} - \tilde \nu_j \, \tilde e_{Li})\, \tilde e_{Lk}^\dagger
\right] + ({\rm h.c.}) \nonumber\\
&&
- \lambda'_{ijk} \left[ 
-m_{u_j} \tilde e_{Li} \tilde d_{Rk}^\dagger \tilde u_{Rj}
-m_{e_i} \tilde u_{Lj} \tilde d_{Rk}^\dagger \tilde e_{Ri}
+m_{d_j} \tilde \nu_i \, \tilde d_{Rk}^\dagger \tilde d_{Rj}
+m_{d_k} (\tilde \nu_i \, \tilde d_{Lj} - \tilde e_{Li}\, \tilde u_{Lj} )\, \tilde d_{Lk}^\dagger
\right] + ({\rm h.c.}) \ ,
\label{eq:rpvlagrangian}
\end{eqnarray}
\end{widetext}
where $P_L \equiv \frac{1}{2} (1-\gamma_5)$.
These RPV interactions are all lepton number violating interactions.
The first two lines give the well-known RPV Yukawa type interactions.
The last two lines are scalar three point interactions which were generated by combining the RPV superpotential and the standard Higgs-matter superpotential of the $R$-parity conserving sector.

\begin{figure}[htbp]
\includegraphics[width=5.2cm]{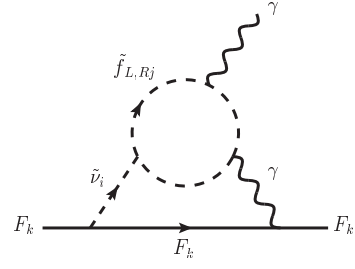}
\caption{\label{fig:sfermion_Barr-Zee} 
Example of Barr-Zee type two-loop contributions to the fermion EDM with sfermion loop in $R$-parity violation.
}
\end{figure}

The EDM of fermion $F$ is defined as follows:
\begin{equation}
{\cal L}_{\rm EDM} = -i  \frac{d_F}{2} \bar \psi \gamma_5 \sigma_{\mu \nu} \psi F^{\mu \nu } \, ,
\end{equation}
and the chromo-EDM of quark $q$ as 
\begin{equation}
{\cal L}_{\rm cEDM} = -i  \frac{d^c_q}{2} g_s \bar \psi \gamma_5 \sigma_{\mu \nu} t_a \psi G^{\mu \nu }_a \, ,
\end{equation}
where $F^{\mu\nu}$ and $G^{\mu\nu}_a$ are respectively the electromagnetic and gluon field strengths.
The RPV lagrangian (\ref{eq:rpvlagrangian}) generates Barr-Zee type two-loop fermion EDM (and chromo-EDM) with sfermion loop as shown in Fig. \ref{fig:sfermion_Barr-Zee}.

\begin{figure}[htbp]
\includegraphics[width=8cm]{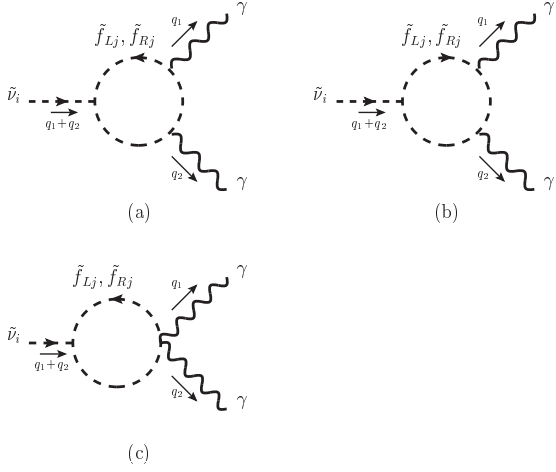}
\caption{\label{fig:nugg_sfermion}
One-loop diagrams contributing to the effective $\tilde \nu \gamma \gamma$ vertex generated by RPV interactions.
}
\end{figure}

The evaluation of the diagrams works as done in previous papers \cite{barr-zee,susyedm2-loop,chang,yamanaka1}. 
We first give the gauge invariant amplitude for the inner sfermion loop effective $\tilde \nu \gamma \gamma$ vertex, and then attach it to the external fermion line.
The one-loop $\tilde \nu \gamma \gamma$ contribution of interest is shown in Fig. \ref{fig:nugg_sfermion}.
It is given as
\begin{widetext}
\begin{eqnarray}
i{\cal M}_{\rm (a)+(b)} &=&
\sum_{\tilde f_j = \tilde f_{Lj},\tilde f_{Rj}}
2 \hat \lambda_{ijj} n_c m_{f_j} (Q_f e)^2 \epsilon^*_\mu (q_1) \epsilon^*_\nu (q_2) 
\int \frac{d^4 k }{(2\pi)^4}
\frac{(2k^\mu + q_1^\mu ) (2k^\nu - q_2^\nu) }{\left[(k+q_1)^2-m_{\tilde f_j}^2 \right] \left[k^2-m_{\tilde f_j}^2\right] \left[(k-q_2)^2-m_{\tilde f_j}^2 \right]} \nonumber\\
&=&
\sum_{\tilde f_j = \tilde f_{Lj},\tilde f_{Rj}}
\hat \lambda_{ijj} n_c m_{f_j} (Q_f e)^2 \epsilon^*_\mu (q_1) \epsilon^*_\nu (q_2)   \nonumber\\
&&\times \Biggl\{
\frac{ 2ig^{\mu \nu}  }{(4\pi )^2} \left[ \frac{2}{\epsilon} -\gamma +\log ( 4\pi ) \right]
-\frac{4 i g^{\mu \nu}}{(4\pi )^2} \int^1_0 \hspace{-.5em} dx \int_0^{1-x} \hspace{-1em} dz\, \log \left( -x(1-x)q_1^2 -z(1-z)q_2^2 -2xz (q_1 \cdot q_2) + m_{\tilde f_j}^2 \right) 
\nonumber\\
&& \hspace{1.5em}
+\frac{2i}{(4\pi)^2}\int^1_0\ dx \int_0^{1-x} \hspace{-.5em}dz\, 
\frac{ - 2x(1- 2x)q_1^\mu q_1^\nu -(1- 2x)(1-2z)q_1^\mu q_2^\nu -4xzq_2^\mu q_1^\nu -2z(1-2z)q_2^\mu q_2^\nu }{ x(1-x)q_1^2 +z(1-z)q_2^2 +2xz(q_1\cdot q_2)-m_{\tilde f_j}^2 }
\Biggr\}
\ ,
\label{eq:sfermionloop(a)(b)}
\\
i{\cal M}_{\rm (c)} &=&
\sum_{\tilde f_j = \tilde f_{Lj},\tilde f_{Rj}}
-2\hat \lambda_{ijj} n_c m_{f_j} (Q_f e)^2 \epsilon^*_\mu (q_1) \epsilon^*_\nu (q_2) 
\int \frac{d^4 k }{(2\pi)^4}
\frac{g^{\mu \nu} }{\left[(k+q_1+q_2)^2-m_{\tilde f_j}^2 \right] \left[k^2-m_{\tilde f_j}^2\right] }
\nonumber\\
&=&
\sum_{\tilde f_j = \tilde f_{Lj},\tilde f_{Rj}}
\hat \lambda_{ijj} n_c m_{f_j} (Q_f e)^2 \epsilon^*_\mu (q_1) \epsilon^*_\nu (q_2) 
\nonumber\\
&&\hspace{4em}
\times \Biggl\{
\frac{-2i }{(4\pi )^2}  \left[ \frac{2}{\epsilon} -\gamma + \log (4\pi) \right]
+\frac{2i }{(4\pi )^2} \int_0^1 dy \log \left( -y(1-y)(q_1+q_2)^2 +m_{\tilde f_j}^2 \right) 
\Biggr\} g^{\mu \nu}
\, ,
\label{eq:sfermionloop(c)}
\end{eqnarray}
\end{widetext}
where diagrams (a) and (b) in  Fig. \ref{fig:nugg_sfermion} were taken together.
Indices $i$ and $j$ denote the flavor of exchanged sneutrino $\tilde \nu$ and loop sfermion $\tilde f$, respectively.
$\hat \lambda $ is the RPV coupling, $\hat \lambda = \lambda $ when the charged slepton runs in the loop, and $\hat \lambda = \lambda'$ in the case of down-type squark. 
The color number $n_c = 1$ ($n_c = 3$ ) if $\tilde f_j$ is a slepton (squark).
$m_{\tilde f_j}$ and $Q_f$ are the mass and the charge in unit of $e$ of the loop sfermion, respectively.
Note that in the above equations, there is also a factor of fermion mass $m_{f_j}$ which issues from the RPV scalar three-point interactions.
The above amplitudes are each divergent, and were treated using dimensional regularization.
The divergence is given by the terms $\frac{2}{\epsilon} -\gamma + \log (4\pi)$, with $\epsilon$ the small shift of the space-time dimension from 4.
We see from the above equations that the divergence cancels in the total contribution $i{\cal M}_{\rm (a)+(b)}+i{\cal M}_{\rm (c)}$, which is one of the consequence of the gauge invariance.

The total sfermion loop Barr-Zee type contribution can be further expanded in terms of the external momentum $q_2$.
Taking up to the first order in $q_2$, we obtain
\begin{eqnarray}
i{\cal M}_{\tilde \nu \gamma \gamma } 
&\approx &
\sum_{\tilde f_j = \tilde f_{Lj},\tilde f_{Rj}}
\frac{i}{(4\pi)^2}\hat \lambda_{ijj} n_c m_{f_j} (Q_f e)^2 \epsilon^*_\mu (q_1) \epsilon^*_\nu (q_2) \nonumber\\
&&
\cdot \left( (q_1 \cdot q_2 )g^{\mu \nu} - q_2^\mu q_1^\nu \right)
\int^1_0\ \hspace{-.5em} dx\, \frac{ 2x (1-x) }{x(1-x)q_1^2 - m_{\tilde f_j}^2}
\, . \nonumber\\
\end{eqnarray}
The above manipulation is suitable, since the EDM is the first order coefficient of the multipole expansion.

\begin{figure}[htb]
\begin{center}
\includegraphics[width=8cm]{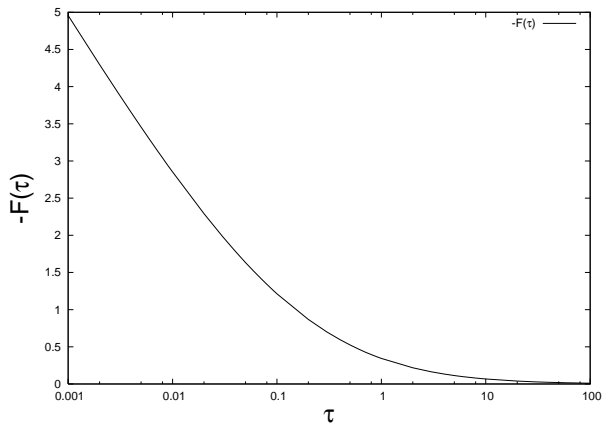}
\caption{\label{fig:bz_integ}
The function $-F(\tau)$ plotted in function of $\tau$.
It is a monotonically decreasing function in $\tau$.
For small $\tau$, $F(\tau) \approx 2+\log(\tau)$.
}
\end{center}
\end{figure}

Now that we have the effective $\tilde \nu \gamma \gamma$ vertex, the remaining part of the computation goes exactly like that of the Barr-Zee type EDM with fermion loop \cite{yamanaka1}.
We insert the effective $\tilde \nu \gamma \gamma$ vertex into the whole Barr-Zee type diagram.
Then we end up with
\begin{widetext}
\begin{eqnarray}
i{\cal M}_{\rm BZ}^{(1)}
&=&
-\frac{1}{2(4\pi)^2} \sum_{\tilde f_j = \tilde f_{Lj},\tilde f_{Rj}}
\hat \lambda_{ijj} \tilde \lambda_{ikk}^* n_c m_{f_j} Q_F e (Q_f e)^2 \epsilon^*_\nu (q)\nonumber\\
&& \hspace{.5em}
\times \int \frac{d^4 k}{(2 \pi)^4} \frac{\bar u (p-q) \gamma_\mu (p\hspace{-.45em}/ \, -q\hspace{-.45em}/ \, -k\hspace{-.4em}/ \, +m_{F_k}) \gamma_5 u(p) \cdot \left( (q\cdot k)g^{\mu \nu} - q^\mu k^\nu \right)}{k^2 \left[ (p-q-k)^2-m_{F_k}^2 \right] \left[ (k+q)^2 - m_{\tilde \nu_i}^2 \right]}
\int_0^1 dx \frac{2x(1-x)}{x(1-x)k^2 -m_{\tilde f_j}^2} \nonumber\\
&\approx &
-\sum_{\tilde f_j = \tilde f_{Lj},\tilde f_{Rj}}
 \hat \lambda_{ijj} \tilde \lambda_{ikk}^* \frac{\alpha_{\rm em}}{2(4\pi)^3} n_c Q_F Q_f^2 e 
\frac{m_{f_j} }{m_{\tilde \nu_i}^2} F(\tau_{\tilde f_j})
\, \epsilon^*_\nu (q) \bar u \sigma^{\mu \nu} q_\mu \gamma_5 u 
\, ,
\label{eq:8}
\end{eqnarray}
\end{widetext}
where $\tilde \lambda = \lambda$ for lepton EDM contribution and $\tilde \lambda = \lambda'$ for quark EDM contribution, and $\tau_{\tilde f_j} \equiv m_{\tilde f_j}^2 / m_{\tilde \nu_i}^2$.
The function $F( \tau)$ is defined as
\begin{eqnarray}
F( \tau) \equiv -\int_0^1 dx
\frac{x(1-x)}{x(1-x) -\tau } \log \left( \frac{x(1-x)}{\tau} \right) \, .
\end{eqnarray}
The shape of the function $F(\tau)$ is depicted in Fig. \ref{fig:bz_integ}.
For small $\tau$, we have $F( \tau) \approx 2+\log \tau$.

To complete the calculation of the Barr-Zee type EDM with sfermion loop, we also have to add diagrams with sneutrino flow inverted and contributions with propagators of exchanged bosons (sneutrino and photon) twisted, and take their imaginary part.
The final result for the EDM is given by
\begin{eqnarray}
d_{F_k}^{\rm sfermion}&=&
-\sum_{\tilde f_j = \tilde f_{Lj},\tilde f_{Rj}}
{\rm Im} (\hat \lambda_{ijj} \tilde \lambda_{ikk}^*) 
\nonumber\\
&&\hspace{4em} \times
\frac{\alpha_{\rm em} n_c Q_F Q_f^2 e}{32\pi^3}  
\frac{m_{f_j} }{m_{\tilde \nu_i}^2} F\bigl( \tau_{\tilde f_j} \, \bigr) \, . \hspace{2em} 
\end{eqnarray}
We show also the result for the chromo-EDM:
\begin{equation}
d_{q_k}^{c;{\rm sfermion}}=
-\sum_{\tilde q_j = \tilde q_{Lj},\tilde q_{Rj}}
{\rm Im} (\lambda'_{ijj} {\lambda'}_{ikk}^*) \frac{\alpha_s}{64\pi^3}  
\frac{m_{q_j} }{m_{\tilde \nu_i}^2} F\bigl( \tau_{\tilde q_j}\, \bigr) \, .
\end{equation}
Let us recall that the Barr-Zee type EDM with fermion loop generated by RPV interactions is given by \cite{yamanaka1}
\begin{equation}
d_{F_k}^{\rm fermion} = {\rm Im} (\hat \lambda_{ijj} \tilde \lambda^*_{ikk}) \frac{\alpha_{\rm em} n_c Q_f^2 Q_F e }{16\pi^3} \frac{m_{f_j}}{m_{\tilde \nu_i}^2} F(\tau_{f_j}) ,
\end{equation}
where $\tau_{f_j} \equiv \frac{m_{f_j}^2}{m_{\tilde \nu_i}^2}$.
We see that the Barr-Zee type EDM with sfermion loop generated by $R$-parity violation has a close form compared with that with fermion loop.
Both involve the same coupling constants, and even the coefficients are the same if $\tilde f_{Lj}$ and $\tilde f_{Rj}$ provide equal contribution.

The difference is given by the overall sign and the loop integral $F(\tau)$.
The opposite sign between fermion and sfermion loop contribution can be explained by the minus sign due to the inner fermion loop of the effective $\tilde \nu \gamma \gamma$ vertex.
This means that the Barr-Zee type RPV EDM with sfermion loop acts destructively to the fermion loop process.
This result is consistent with the analyses of the one-loop level decay of the Higgs boson into two gauge bosons (photons or gluons), where we also encounter the interference of fermion and sfermion loops  \cite{higgsdecay} (in these analyses, only Higgs-matter interactions are involved, but the generation of Yukawa and scalar three-point interactions from the superpotential goes in the same manner as for the RPV interactions).

To compare the relative size between fermion and sfermion loop contributions, we have just to see the difference between $F\bigl( \tau_{ f_j} \, \bigr)$ and $F\bigl( \tau_{\tilde f_j} \, \bigr)$.
For sparticle masses $m_{\rm SUSY} =1$ TeV, $F\bigl( \tau_{ f_j} \, \bigr)$ is around $-10$ for fermion loop process.
If we assume that $m_{\tilde f_j} \approx m_{\tilde \nu_i}$, i.e. $\tau_{\tilde f_j}\approx 1$, we obtain $F\bigl(\tau_{\tilde f_j} \,  \bigr) \approx -0.34$ (see Fig. \ref{fig:bz_integ}), so sfermion loop diagrams can be neglected.
The situation changes however when the sneutrino mass $m_{\tilde \nu_i}$ is more than 10 times heavier than the loop sfermion mass $m_{\tilde f_j}$ ($\tau_{\tilde f_j} < 10^{-2} $), where $-F(\tau)$ can exceed 3, so that the total RPV Barr-Zee type contribution can be significantly suppressed.
If we take the sum of fermion and sfermion loop processes, we obtain
\begin{equation}
d_{F_k}^{f+\tilde f} \approx {\rm Im} (\hat \lambda_{ijj} \tilde \lambda^*_{ikk}) \frac{\alpha_{\rm em} n_c Q_f^2 Q_F e }{16\pi^3} \frac{m_{f_j}}{m_{\tilde \nu_i}^2} \log \frac{m_{f_j}^2}{m_{\tilde f_j}^2} ,
\end{equation}
for $m_{f_j}^2<<m_{\tilde f_j}^2 << m_{\tilde \nu_i}^2$, where we have assumed that $\tilde f_{Lj}$ and $\tilde f_{Rj}$ provide equal contribution.
We see that it is not possible to cancel completely the RPV Barr-Zee type EDM, as $m_{f_j}^2<<m_{\tilde f_j}^2$, which is known from the current lower limit on sparticle masses \cite{susylhc}, and the fermion loop contribution keeps the largest part.

In conclusion, we have found and analyzed the RPV supersymmetric contribution to the Barr-Zee type two-loop diagrams with sfermion loops.
Our result says that the sfermion loop contribution interferes destructively with the known fermion loop RPV Barr-Zee type effect.
The former can be neglected if all sparticle masses are taken at the same order of magnitude, but can be significant when the sneutrino is much heavier than the sfermion in the loop, and consequently the suppression of the total RPV two-loop level EDM can occur.
It is however not possible to completely cancel the RPV contribution to the EDM with the fermion and sfermion loop diagrams, the latter being smaller than the former.

\begin{acknowledgments}
The author thanks M. Valverde for useful discussions and comments.
\end{acknowledgments}

\end{document}